\documentclass{iaus}
\usepackage{graphicx}

\title[Impact of stellar duplicity on the frequency of giant planets] 
{Probing the impact of stellar duplicity on the frequency of giant planets: 
final results of our VLT/NACO survey}

\author[Anne Eggenberger et al.]   
{Anne Eggenberger$^1$, 
 St\'ephane~Udry$^2$, 
Ga\"el~Chauvin$^1$, 
Thierry~Forveille$^1$,
Jean-Luc~Beuzit$^1$,
Anne-Marie~Lagrange$^1$,
\and Michel Mayor$^2$}

\affiliation{$^1$ Universit\'e Joseph Fourier -- Grenoble 1 / CNRS,  
              Laboratoire d'Astrophysique de Grenoble (UMR 5571), 
	      BP 53, 38041 Grenoble Cedex 9, France \\ 
              email: {\tt Anne.Eggenberger@obs.ujf-grenoble.fr, 
	      Gael.Chauvin@obs.ujf-grenoble.fr,
	      Thierry.Forveille@obs.ujf-grenoble.fr,
	      Jean-Luc.Beuzit@obs.ujf-grenoble.fr,
	      Anne-Marie.Lagrange@obs.ujf-grenoble.fr} \\[\affilskip]
              $^2$ Observatoire de Gen\`eve, Universit\'e de Gen\`eve, 
	      51 ch. des Maillettes, 1290 Sauverny, Switzerland
 \\email: {\tt Stephane.Udry@unige.ch, Michel.Mayor@unige.ch}}

\pubyear{2008}
\volume{xxx}  
\pagerange{119--126}
\jname{Title of your IAU Symposium}
\editors{A.C. Editor, B.D. Editor \& C.E. Editor, eds.}
\begin{document}

\maketitle

\begin{abstract}
If it is commonly agreed that the presence of a (moderately) close stellar 
companion affects the formation and the dynamical evolution of giant 
planets, the frequency of giant planets residing in binary systems separated 
by less than 100 AU is unknown. To address this issue, we have conducted 
with VLT/NACO a systematic adaptive optics search for moderately close 
stellar companions to 130 nearby solar-type stars. According to the data 
from Doppler surveys, half of our targets host at least one planetary 
companion, while the other half show no evidence for short-period giant 
planets. We present here the final results of our survey, which include a 
new series of second-epoch measurements to test for common proper motion. 
The new observations confirm the physical association of two companion 
candidates and prove the unbound status of many others. These results 
strengthen our former conclusion that circumstellar giant planets are 
slightly less frequent in binaries with mean semimajor axes between 35 and 
100 AU than in wider systems or around single stars.
\keywords{planetary systems: formation, binaries: visual, techniques: high angular resolution}
\end{abstract}

\firstsection 

\section{The NACO survey}
To probe the impact of stellar duplicity on the frequency of giant planets, we have conducted with VLT/NACO an adaptive optics search for stellar companions to $\sim$60 planet-host stars and to $\sim$70 non-planet-host stars (hereafter control stars). This survey revealed 95 companion candidates near 33 targets (Fig.~\ref{fig1}). Using two-epoch astrometry we identified 19 true companions, 2 likely bound objects, and 34 background stars (\cite[Eggenberger et al. 2007]{Eggenberger07}). The companionship of the remaining 40 companion candidates could not be constrained due to the lack of a second-epoch astrometric measurement. Assuming that all but two of these 40 objects were unbound, we showed that giant planets seem slightly less frequent in $\sim$35-100 AU binaries than around single stars (\cite[Eggenberger et al. 2008]{Eggenberger08}).

\section{New observations}
We recently performed the second-epoch measurements that were missing previously. The new observations show that the companion candidates we detected near the planet-host stars HD 76700, HD 83443, HD 162020,  and HD 330075 are all unrelated background stars. On the other hand, the new data confirm the physical association of the companion candidates to the control stars HD 82241 and HD 134180.

\section{Statistical analysis}
\begin{figure}[t]
\centering
\resizebox{\textwidth}{!}{
   \includegraphics{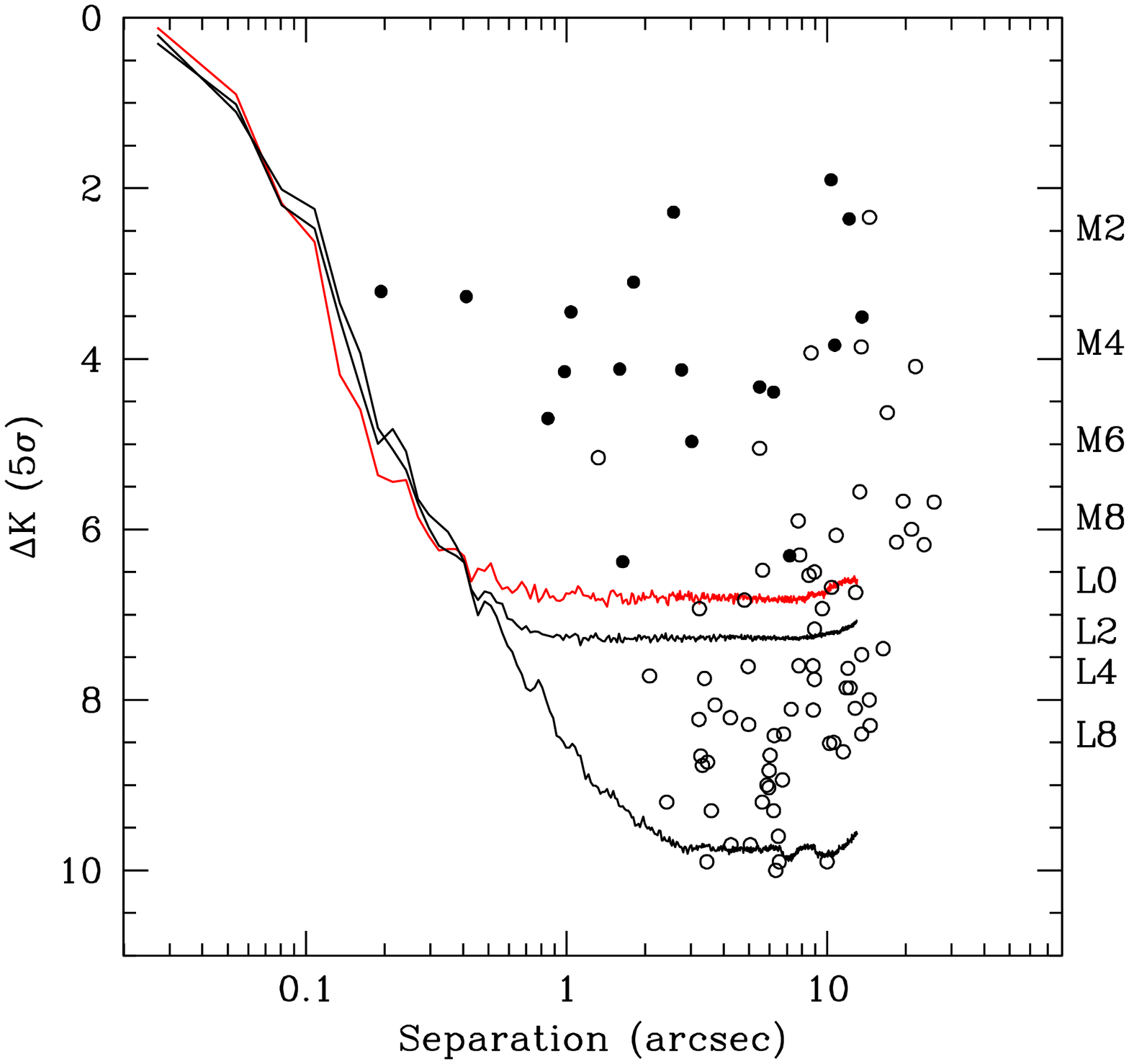}\hspace{0.5cm}
   \includegraphics{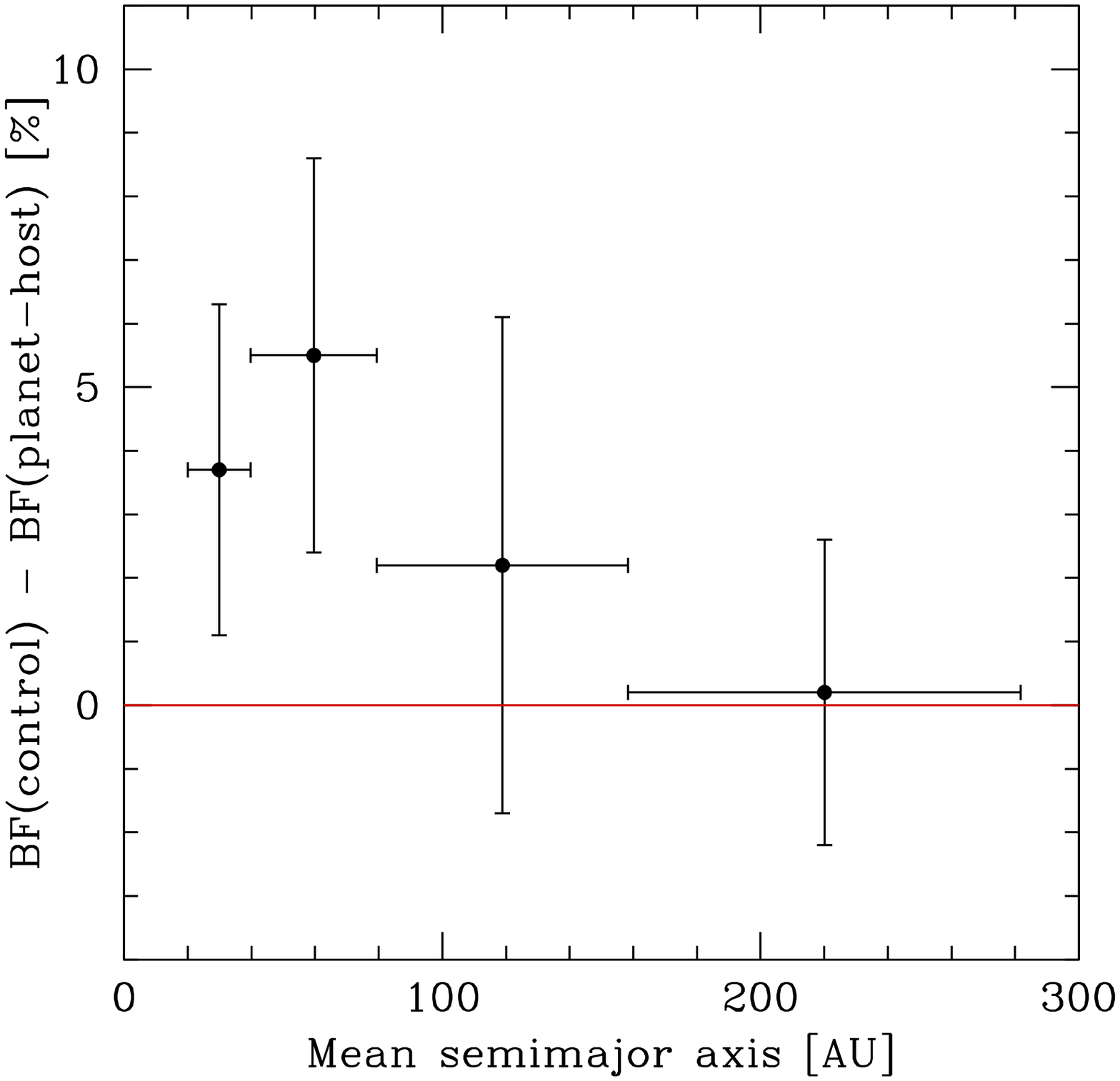} 
}
\caption{Results of our survey. {\it Left:} Detections and detection limits. Dots are true companions, circles are unbound objects. The uppermost curve is the detection limit used for the statistical analysis. The two lower curves are median detection limits obtained with the old and with the present detectors of NACO. {\it Right:} Difference between the binary fraction of control stars and the binary fraction of planet-host stars as a function of binary mean semimajor axis. Vertical error bars are 68\% bootstrap confidence intervals. Horizontal error bars represent the bin width.}
\label{fig1}
\end{figure}

Figure~\ref{fig1} (right) shows the difference in binary fraction between the control and the planet-host subsamples. According to the updated statistical analysis, the difference in binary fraction is $13.2\pm 5.1$\% for semimajor axis below 100 AU, and $-1.5 \pm 2.9$ for semimajor axis between 100 and 200 AU. The positive difference seen for mean semimajor axis $\lesssim$100 AU suggests that giant and intermediate-mass planets are slightly less frequent in moderately close binaries than in wider systems or around single stars. If confirmed with a larger sample, this result would support the idea that the presence of a moderately close stellar companion affects the formation of giant planets, but does not completely stop the process.

\begin{acknowledgements}
AE acknowledges support from the Swiss National Science Foundation through 
a fellowship for advanced researchers. 
\end{acknowledgements}



\end{document}